\documentclass[aip,reprint,showpacs,amsmath,amssymb,a4paper,superscriptaddress,10pt]{revtex4-1}

\usepackage{graphicx}			
\usepackage{bm}           
\usepackage{natbib}       

\usepackage{ifpdf}
\ifpdf
  \usepackage[pdftex]{color}
  \usepackage[pdftex,hyperfootnotes=false,colorlinks=false]{hyperref}
  \hypersetup{
    colorlinks=false,
    pdfpagemode=UseThumbs,
    plainpages=false   }
\else
  \usepackage[dvips]{color}
  \usepackage[dvips,hyperfootnotes=false,colorlinks=false]{hyperref}
  \hypersetup{
    colorlinks=false,
		pdfpagemode=UseThumbs,
    plainpages=false
   }
\fi

\begin{document}

\title{Few electron double quantum dot in an isotopically purified $^{\mathrm{28}}$Si quantum well}

\author{A.~\surname{Wild}}
\affiliation{Walter Schottky Institut and Physics Department, Technische Universit\"at M\"unchen, Am Coulombwall 4, 85748, Garching, Germany}

\author{J.~\surname{Kierig}}
\affiliation{Institut f\"ur Experimentelle und Angewandte Physik, Universit\"at Regensburg, Universit\"atsstra\ss e 31, 93040, Regensburg, Germany}

\author{J.~\surname{Sailer}}
\altaffiliation[Present address: ]{Department of Applied Physics, The University of Tokyo, 7-3-1 Hongo, Bunkyo-ku, Tokyo 113-8656, Japan}
\affiliation{Walter Schottky Institut and Physics Department, Technische Universit\"at M\"unchen, Am Coulombwall 4, 85748, Garching, Germany} 

\author{J.~W.~\surname{Ager III}}
\affiliation{Lawrence Berkeley National Laboratory, Materials Sciences Division, Berkeley, CA 94720-8197, USA}

\author{E.~E.~\surname{Haller}}
\affiliation{Lawrence Berkeley National Laboratory, Materials Sciences Division, Berkeley, CA 94720-8197, USA}
\affiliation{Department of Materials Science and Engineering, University of California at Berkeley, Berkeley, CA 94720-1760, USA}

\author{G.~\surname{Abstreiter}}
\affiliation{Walter Schottky Institut and Physics Department, Technische Universit\"at M\"unchen, Am Coulombwall 4, 85748, Garching, Germany}
\affiliation{Technische Universit\"at M\"unchen Institute for Advanced Study, Lichtenbergstra\ss e 2a, 85748, Garching, Germany}

\author{S.~\surname{Ludwig}}
\affiliation{Fakult\"at f\"ur Physik and Center for NanoScience, Ludwig-Maximilians-Universit\"at M\"unchen, Geschwister-Scholl-Platz 1, 80539, Munich, Germany}

\author{D.~\surname{Bougeard}}
\email{dominique.bougeard@ur.de}
\affiliation{Institut f\"ur Experimentelle und Angewandte Physik, Universit\"at Regensburg, Universit\"atsstra\ss e 31, 93040, Regensburg, Germany}

\date{\today}

\begin{abstract}

We present a few electron double quantum dot device defined in an isotopically purified $^{\mathrm{28}}$Si quantum well (QW). An electron mobility of $5.5 \cdot 10^4~\mathrm{cm}^2(\mathrm{Vs})^{-1}$ is observed in the QW which is the highest mobility ever reported for a two-dimensional electron system in $^{\mathrm{28}}$Si. The residual concentration of $^{\mathrm{29}}$Si nuclei in the $^{\mathrm{28}}$Si QW is lower than $10^{3}~\mathrm{ppm}$, at the verge where the hyperfine interaction is theoretically no longer expected to dominantly limit the $T_{\mathrm{2}}$ spin dephasing time. We also demonstrate a complete suppression of hysteretic gate behavior and charge noise using a negatively biased global top gate. 
\end{abstract}


\maketitle 


Semiconductor quantum dots (QD) are among the candidates for a scalable implementation of electron spin based qubits in solid state systems. Silicon (Si) has been widely recognized as a well suited material system for decoupling electron spin qubits from their volatile solid state environment owing to the weak spin-orbit and weak hyperfine interaction.
Very long spin relaxation times (T$_{\mathrm{1}}$) on the order of seconds have been reported for Si on the basis of electrostatically defined QDs~\cite{Simmons2011}, single phosphorous donors~\cite{Morello2010} or triplet-singlet relaxation times in double QDs~\cite{Prance2011}. Recently, also a spin dephasing time of $T_{\mathrm{2}}^*=360~\mathrm{ns}$ has been observed in a time ensemble measurement in a Si double QD~\cite{Maune2012}. These milestones highlight the great potential for quantum information processing in Si.

The adverse impact from nuclear spins on electron spin coherence~\cite{Petta2005} can be further reduced in the Si material system by means of isotopic enrichment of the $^{\mathrm{28}}$Si isotope which has zero nuclear spin.
Recent technological advances have enabled the fabrication of highly enriched $^{\mathrm{28}}$Si crystals~\cite{Andreas2011} with isotopic fractions of the nuclear spin carrying $^{\mathrm{29}}$Si isotope smaller than $4\cdot10^{2}~\mathrm{ppm}$. In such ultra-clean $^{\mathrm{28}}$Si bulk samples, the spin coherence time $T_{\mathrm{2}}$ for donor-bound electrons~\cite{Tyryshkin2011} achieves unprecedentedly long values of $T_{\mathrm{2}}=10~\mathrm{s}$. This offers a promising perspective for qubit applications with electrostatically defined QDs in $^{\mathrm{28}}$Si heterostructures.
However, the integration of isotopically purified material with low impurity concentrations into molecular beam epitaxy (MBE) or chemical vapor deposition growth processes is still a challenge. Hence, no QD devices have been demonstrated so far for two-dimensional electron systems (2DES) in $^{\mathrm{28}}$Si.

In this letter, we report on the fabrication and characterization of an electrostatically defined few electron double QD within a high mobility 2DES in a MBE-grown $^{\mathrm{28}}$Si/SiGe heterostructure. We find a concentration of residual $^{\mathrm{29}}$Si nuclei in the  quantum well (QW) smaller than $10^3~\mathrm{ppm}$ and achieve a peak mobility of $5.5\cdot10^4~\mathrm{cm}^2(\mathrm{Vs})^{-1}$ at a 2DES density of $3\cdot10^{11}~\mathrm{cm}^{-2}$. We combine our double QD with a global top gate (TG) and demonstrate a strong suppression of hysteretic gate behavior and charge noise as a negative voltage is applied to the global TG.

\begin{figure}	\centering
		\includegraphics[width=1.00\columnwidth]{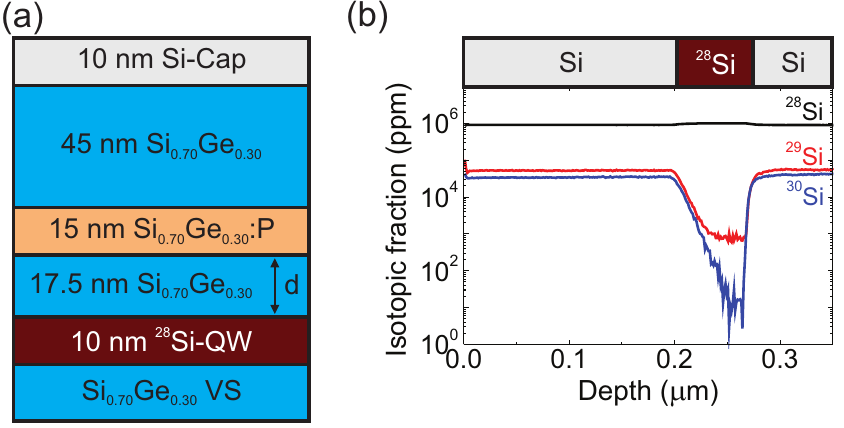}
	\caption{\label{Fig1} (Color Online) (a) Layer sequence of the $^{\mathrm{28}}$Si/SiGe heterostructure. A $^{\mathrm{28}}$Si QW is embedded into a natural SiGe host crystal and separated by a spacer of thickness $d$ from the SiGe:P layer. (b) SIMS measurement of the $^{\mathrm{28,29,30}}$Si isotopic fractions in a test structure (top).}
\end{figure}

Our heterostructures are grown in a solid source MBE system equipped with independent electron beam evaporators for Si and Ge of natural isotopic composition as well as $^{\mathrm{28}}$Si. All evaporators are equipped with high purity single crystals as source materials. 
The SiGe heterostructure discussed here is sketched in Fig.~\ref{Fig1}(a) and contains three key elements. The first is a relaxed SiGe virtual substrate (VS) grown by using Si and Ge of natural isotopic composition. The second is a $10~\mathrm{nm}$ thick $^{\mathrm{28}}$Si QW hosting the 2DES. The third element is the modulation doping. It consists of a SiGe:P layer with a phosphorus (P) concentration of $10^{18}~\mathrm{cm}^{-3}$ and a spacer layer of thickness $d=17.5~\mathrm{nm}$ which separates the QW from the remote dopants.

To verify the chemical purity of our isotopically enriched MBE grown material, we employ high resolution secondary ion mass spectrometry (SIMS). We found no contamination of the $^{\mathrm{28}}$Si layers by spurious elements 
compared to the intrinsic substrate and to typical structures grown from our source material of natural isotopic composition. This rules out potential contaminations during the preparation of the isotopically enriched $^{\mathrm{28}}$Si MBE source crystal. Furthermore, we determined the residual concentration of $^{\mathrm{29}}$Si in our epitaxial $^{\mathrm{28}}$Si. To enable a high accuracy concentration measurement, a test structure as sketched at the top of Fig.~\ref{Fig1}(b) was grown. A $75~\mathrm{nm}$ thick layer of $^{\mathrm{28}}$Si is sandwiched between two natural Si regions~\cite{Note1}. Within the $^{\mathrm{28}}$Si layer, the residual concentration of the nuclear spin carrying $^{\mathrm{29}}$Si isotope drops below an isotopic fraction of $10^{3}~\mathrm{ppm}$. This concentration can be assessed with the help of recent theoretical studies. By investigating decoherence in isotopically enriched $^{\mathrm{28}}$Si:P structures, Witzel~et~\textit{al.}~\cite{Witzel2010} found that the spin coherence time $T_{\mathrm{2}}$ is no longer solely limited by the hyperfine interaction but increasingly impaired by dipolar interactions with paramagnetic impurities - which are unavoidably present in any real crystal - below a threshold of roughly $10^{3}~\mathrm{ppm}$ for $^{\mathrm{29}}$Si. Furthermore, Assali~et~\textit{al.}~\cite{Assali2011} calculated the hyperfine induced dephasing of an electron spin for gate-defined QDs in Si for various $^{\mathrm{29}}$Si concentrations which is in good agreement with recent experimental data~\cite{Maune2012}. Interpolating their numerical results, the model predicts a spin dephasing time on the order of $T_{\mathrm{2}}^*\approx2~\mathrm{\mu s}$ for the $^{\mathrm{29}}$Si concentration of $10^{3}~\mathrm{ppm}$ in our material~\cite{Note3} which would represent a strong improvement over $T_{\mathrm{2}}^*\approx10~\mathrm{ns}$ measured in GaAs~\cite{Petta2005}. 

\begin{figure}
	\centering
		\includegraphics[width=1.00\columnwidth]{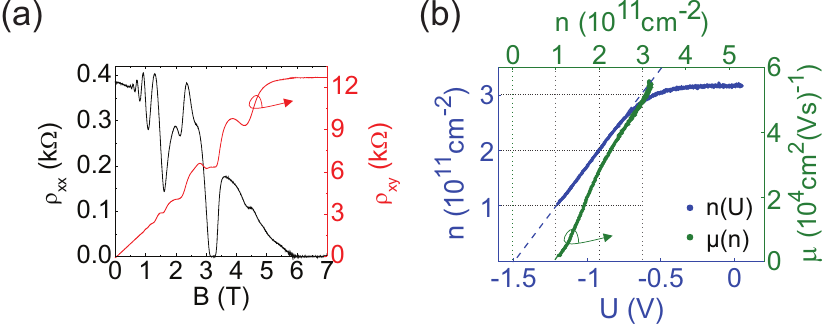}
	\caption{	\label{Fig2} (Color Online) Magneto-transport measurements on a Hall bar obtained at $320~\mathrm{mK}$ (a) Longitudinal ($\rho_{\mathrm{xx}}$) and transversal ($\rho_{\mathrm{xy}}$) resistivity as a function of perpendicular magnetic field B of the 2DES in the $^{\mathrm{28}}$Si QW. (b) 2DES density $n$ as a function of gate voltage (left/bottom axis) and mobility $\mu$ as a function of density (right/top axis) for a Hall bar with an Al$_{\mathrm{2}}$O$_{\mathrm{3}}$ gate dielectric. The density was determined from the low field slope of $\rho_{\mathrm{xy}}$.}
\end{figure}

Using the above heterostructure, we followed a recipe published earlier~\cite{Sailer2009,Sailer2010} to fabricate Hall bar devices. Figure~\ref{Fig2}(a) shows results from a magneto-transport measurement. The observation of Shubnikov-de Haas (SdH) oscillations and well resolved integer quantum Hall plateaus identifies the existence of a high quality 2DES in the $^{\mathrm{28}}$Si layer. 

To manipulate the 2DES density, the Hall bar is covered by a Palladium (Pd) gate on top of a $20~\mathrm{nm}$ thick Al$_{\mathrm{2}}$O$_{\mathrm{3}}$ dielectric which is fabricated by means of atomic layer deposition (ALD). 
The 2DES density $n$ is shown in Fig.~\ref{Fig2}(b) as a function of gate voltage $U$ and can be tuned between $0$ and $3\cdot10^{11}~\mathrm{cm}^{-2}$. 
Above $U=-0.7~\mathrm{V}$, $n$ is almost independent of $U$, while below $U=-0.7~\mathrm{V}$, $n$ depends linearly on gate voltage.
The extrapolation of $n(U)$ [dashed line in Fig.~\ref{Fig2}(b)] implies that the 2DES is completely depleted at $U=-1.5~\mathrm{V}$.

The measured 2DES mobility $\mu(n)$ is plotted in the same graph. Starting from zero mobility at a finite density of $n_{\mathrm{min}} \approx 1\cdot10^{11}~\mathrm{cm}^{-2}$, the data exhibits a peak mobility of $5.5\cdot10^4~\mathrm{cm}^2(\mathrm{Vs})^{-1}$ at a density of about $3\cdot10^{11}~\mathrm{cm}^{-2}$. This is the highest mobility ever reported for a 2DES in $^{\mathrm{28}}$Si~\cite{Sailer2009,Shankar2010,Lo2011}. 
Control experiments, using heterostructures equivalent to the one in Fig.~\ref{Fig1}(a) but with different spacer thicknesses $d$, reveal a strong superlinear dependence of the mobility on $d$ (not shown). 
This points towards a notable effect of remote impurities located above the QW on the mobility in the 2DES. The particular influence of impurities introduced by the modulation doping layer on the electron mobility has been assessed through detailed calculations by A.~Gold~\cite{Gold1991,Gold2010}. These calculations were compared in a wide density range to experimental data from a Si/SiGe heterostructure almost identical to ours~\cite{Gold2010}. The mobility is found to be determined by remote dopants in the SiGe:P layer. In the regime of low densities, the model predicts the dopants to create strong disorder which likely induces an Anderson-type metal insulator transition (MIT) at a density $N_{\mathrm{MIT}}$. For our system, we find good agreement between the model and our data in the whole covered density range~\cite{Note2}. 
Especially $N_{\mathrm{MIT}}=0.95 \cdot 10^{11}~\mathrm{cm}^{-2}$ can be calculated (Eq.~6 in A.~Gold~\cite{Gold1991}) and matches nicely with $n_{\mathrm{min}}$ where the mobility drops to zero.
From the strong $d$ dependence of $\mu$ as well as the agreement of $\mu(n)$ with theory, we conclude that remote dopants in the SiGe:P layer constitute the main mobility limiting mechanism rather than impurities in the $^{\mathrm{28}}$Si QW. These results support our SIMS analysis and the high chemical purity of the MBE grown isotopically enriched material.
 
\begin{figure}
	\centering
		\includegraphics[width=1.00\columnwidth]{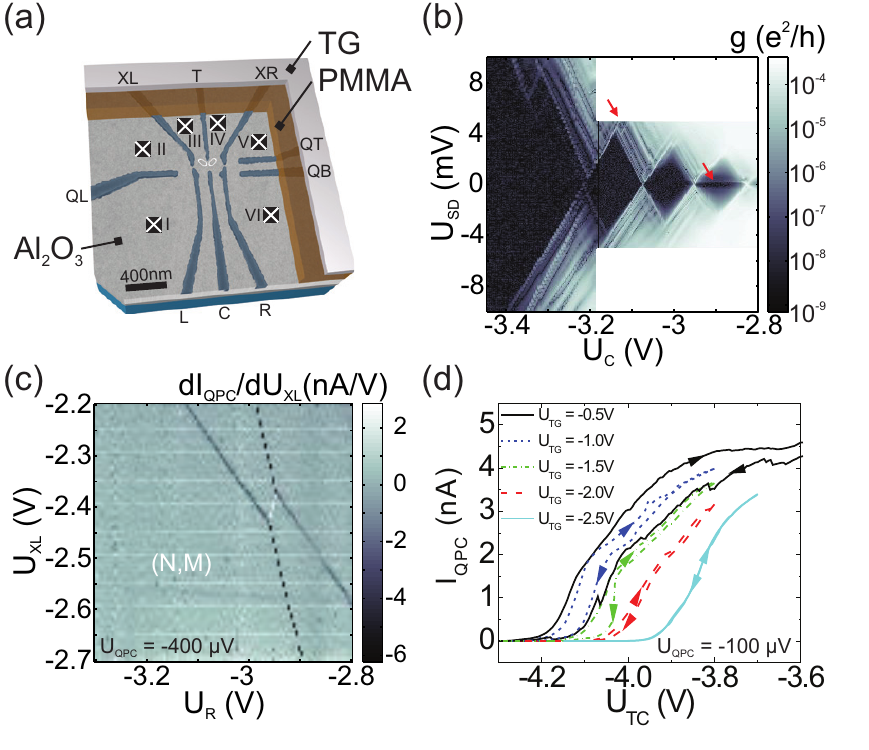}
	\caption{	\label{Fig3} (Color Online) (a) Layer stack of the device and electron beam defined gate layout. The double QD is sketched with white ellipses. Roman numbers denote ohmic contacts. (b) Coulomb diamond measurement of a single few electron QD showing the differential conductance $g$ as a function of gate ($U_{\mathrm{C}}$) and source-drain ($U_{\mathrm{SD}}$) voltage. (c) Charge stability diagram of the double QD showing the transconductance $\partial I_{\mathrm{QPC}}/\partial U_{\mathrm{XL}}(U_{\mathrm{XL}},U_{\mathrm{R}})$. (d) Current $I_{\mathrm{QPC}}$ flowing from contact III to IV as a function of the voltage $U_{\mathrm{TC}}$ applied to gates T and C for different global top gate voltages $U_{\mathrm{TG}}$. The gate sweep direction is indicated by arrows.
The cryostat base temperature was $T \simeq 20~\mathrm{mK}$ in (b), (c), (d).}
\end{figure}

The layer stack and the gate layout of our double QD device is shown in Fig.~\ref{Fig3}(a). The QD gates are separated from the Si surface by $20~\mathrm{nm}$ of Al$_{\mathrm{2}}$O$_{\mathrm{3}}$ in order to minimize the risk of leakage currents~\cite{Dotsch2001,Yuan2011}. Above the gate layer, we introduce a film of cross-linked PMMA with a thickness of $140~\mathrm{nm}$ which serves as a gate insulator for an additional global Pd TG. 

First, we tune the global TG to $U_{\mathrm{TG}}=-4~\mathrm{V}$ and apply negative voltages to gates T, XR, R and C to form a single QD while all other gates are grounded. Figure~\ref{Fig3}(b) presents the differential conductance of this QD as a function of $U_{\mathrm{C}}$. The Coulomb diamond (CD) sizes increase with more negative $U_{\mathrm{C}}$ from which charging energies $1.5~\mathrm{meV}\leqq E_{\mathrm{C}}\leqq10~\mathrm{meV}$ can be extracted. Such a strong dependence of $E_{\mathrm{C}}(U_{\mathrm{C}})$ indicates that the QD is already in the few electron regime. Arrows mark co-tunneling features and a rich spectrum of excited states which are evidenced by the existence of many conductance lines parallel to the CD edges. From these features, electronic excitation energies of $\approx 250~\mathrm{\mu eV}$ can be extracted which is consistent with other QDs in Si/SiGe~\cite{Wild2010,Thalakulam2010}. The CDs also allow to determine the capacitive coupling $C_g=\alpha_ge^2E^{-1}_{\mathrm{C}}$ between the QD and its gates and the corresponding conversion factors $\alpha_{g}$~\cite{Ihn2010} to $\alpha_{\mathrm{C}}=0.0246$, $\alpha_{\mathrm{T}}=0.0365$ and $\alpha_{\mathrm{XR}}=0.0261$. The relative sizes of these lever arms suggest a QD position as sketched by the right white circle in Fig.~\ref{Fig3}(a).

Next, we form a double QD also at $U_{\mathrm{TG}}=-4~\mathrm{V}$ by additionally energizing the left gates of the device [Fig.~\ref{Fig3}(a)]. In addition, we bias gates QT and QB to define a quantum point contact (QPC) as a charge sensor. Figure~\ref{Fig3}(c) shows the charge stability diagram of the double QD illustrated by the transconductance $\partial I_{\mathrm{QPC}}/\partial U_{\mathrm{XL}}$ as a function of the voltages applied to gates XL and R. Dark lines with negative slope represent charging lines of the double QD in the few electron regime while the white line with positive slope corresponds to an inter-dot transition. 
$(N,M)$ labels the last charging state of the double QD we can detect. For higher gate voltages, the occupation of the double QD increases by one electron in each QD. Beyond, the double well potential transforms into a soft single well as the inter-dot barrier decreases strongly with plunger gate voltages.

We finally evaluate the benefit of the global TG on device performance. Therefore, we define a QPC via gates T and C and measure $I_{\mathrm{QPC}}$ as a figure of merit for the stability of the local potential in the vicinity of the double QD. Figure~\ref{Fig3}(d) shows pinch-off curves of $I_{\mathrm{QPC}}$ as a function of $U_{\mathrm{TC}}$ for five distinct voltages $U_{\mathrm{TG}}$ where arrows mark up and down sweeps of $U_{\mathrm{TC}}$. 
For $U_{\mathrm{TG}}\geqq-1.5~\mathrm{V}$, we observe a strongly hysteretic behavior of $I_{\mathrm{QPC}}(U_{\mathrm{TC}})$ for successive up and down sweeps of $U_{\mathrm{TC}}$ as well as random, abrupt switching events in $I_{\mathrm{QPC}}$. Such switching noise has impaired measurements before in SiGe QDs~\cite{Wild2010,Payette2011}.
Around $U_{\mathrm{TG}}=-2~\mathrm{V}$, gate hysteresis and switching events are less pronounced, whereas both features vanish below $U_{\mathrm{TG}}\leqq-2.5~\mathrm{V}$ as shown in the rightmost trace of Fig~\ref{Fig3}(d).  
Similarly beneficial effects of a global TG on the suppression of switching events for a QPC in GaAs have been observed by Buizert and colleagues~\cite{Buizert2008}. They related the occurrence of switching noise to gate-2DES leakage currents that can be suppressed by increasing the effective height of the barrier at the surface for tunneling electrons via a negatively biased global TG. Another work on bias cooling applied to GaAs QDs points in the same direction~\cite{Pioro-Ladriere2005}.
In contrast, we exclude tunneling processes from any gate at the surface into the heterostructure as the origin of switching noise in our device. Due to the Al$_{\mathrm{2}}$O$_{\mathrm{3}}$ insulator, we can apply up to $-12~\mathrm{V}$ between the QD gates and the 2DES without leakage. 
Since biasing a global TG modifies the band structure between the global TG and the 2DES, the gradual suppression of switching noise and gate hysteresis with more negative values of $U_{\mathrm{TG}}$ indicates that the global TG acts on charge traps located between the 2DES and the sample surface. Thus, we suggest that a global TG voltage of $U_{\mathrm{TG}}=-4~\mathrm{V}$ either depletes charge traps or localizes fluctuating charges in long-lived states and, as a result, enables the stable operation of our double QD.
Even beyond issues related to obvious device stability, a global TG could also turn out advantageous for qubit operation in isotopically enriched $^{\mathrm{28}}$Si. On the one hand, it can reduce the number of trapped spins that induce decoherence of a spin qubit~\cite{Witzel2010} through depletion. On the other hand, charge fluctuations could be frozen out, which otherwise degrade the qubit coherence time via the exchange, spin-orbit or hyperfine interaction.

In summary, we presented a few electron double QD device in a nuclear spin refined $^{\mathrm{28}}$Si QW with a residual concentration of nuclear spin carrying $^{\mathrm{29}}$Si nuclei smaller than $10^{3}~\mathrm{ppm}$. For this concentration, a recent theory predicts an almost three orders of magnitude increase of the spin dephasing time compared to GaAs~\cite{Assali2011}. The 2DES achieves a record mobility of $5.5 \cdot 10^4~\mathrm{cm}^2(\mathrm{Vs})^{-1}$ which is only limited by remote impurities in the doping layer. Hence, there is no intrinsic limitation for reaching lower levels of disorder in the 2DES from the usage of the isotopically purified source material. We discussed the beneficial role a global TG can adopt within the device regarding stability and coherence of qubit states.
Altogether, our findings render isotopically purified $^{\mathrm{28}}$Si/SiGe heterostructures an interesting platform for future applications in quantum information. \\

This work was supported by the Deutsche Forschungsgemeinschaft via SFB~631 and the "Nano Initiative Munich" (NIM). We gratefully thank Daniela Taubert, Gunnar Petersen and Rupert Huber for technical assistance and Dirk Grundler for access to the Al$_{\mathrm{2}}$O$_{\mathrm{3}}$ ALD funded via the European Community's FP7/2007-2013 program under grant No.~228673. Work at the LBNL was supported in part by US NSF Grant No.~DMR-0405472 and the U.S. DOE under Contract No.~DE-AC02-05CH11231.

\end{document}